\documentstyle[12pt,epsfig]{article}
\newcommand{\beq}{\begin{equation}}
\newcommand{\eeq}{\end{equation}}
\newcommand{\beqs}{\begin{eqnarray}}
\newcommand{\eeqs}{\end{eqnarray}}
\newcommand{\ce}{c_{\mathrm eff.}}
\newcommand{\indi}{\nu_1\cdots\nu_{p+1}}
\newcommand{\tym}{{{\cal T}_{\mathrm YM}}}
\newcommand{\indii}{\nu_1\cdots\nu_{p+2}}
\begin{document}
\begin{titlepage}
\begin{flushleft}
       \hfill                      {\tt hep-th/9811208}\\
       \hfill                       SISSA ref. 127/98/EP\\
\end{flushleft}
\vspace*{3mm}
\begin{center}
{\LARGE On the construction of gauge theories\\ }
\vspace*{5mm}
{\LARGE from non critical type 0 strings\\ }
\vspace*{12mm}
{\large Gabriele Ferretti\footnote{\tt ferretti@sissa.it}}\\
\vspace*{2mm}
and\\
\vspace*{2mm}
{\large Dario Martelli\footnote{\tt dmartell@sissa.it}}\\
\vspace*{4mm}
{\it SISSA, Via Beirut 2 Trieste 34014, Italy\\}
{\it and\\}
{\it INFN, Sezione di Trieste\\}
\vspace*{10mm}
\end{center}

\begin{abstract}

We investigate Polyakov's proposal of constructing Yang-Mills theories by
using non critical type 0 strings. 
We break conformal invariance by putting the system
at finite temperature
and find that the entropy of the cosmological solutions for these theories 
matches that of a gas of weakly interacting Yang-Mills bosons, up to a 
numerical constant. The computation of the entropy using the effective 
action approach presents some novelties in that the whole contribution 
comes from the RR fields. We also find an area law and a mass gap in the 
theory and  show that such behavior persists for $p>4$. We comment on the
possible physical meaning of this result.

\end{abstract}
\end{titlepage}

\section{Introduction}

The recent progress in string theory has unearthed deep connections
between gravity and gauge theories. The conjecture \cite{M1,GKP1,W1} 
that ${\cal N}=4$ super Yang-Mills theory is dual to type IIB supergravity
in $AdS_5\times S^5$ has led to the computation of many correlation functions
of local observables and Wilson loops for this theory.
An even more exciting 
possibility is the applicability of these techniques to non supersymmetric 
gauge theories. In this context, Witten \cite{W2} has 
proposed to compactify the $4+1$ dim.
theory of a D4-brane on a circle with supersymmetry breaking boundary
conditions, leaving a $3+1$ dimensional pure Yang-Mills theory with
an effective cut-off given by the radius of the circle. This proposal has
enjoyed many successes, yielding qualitatively reasonable results for
bare quantities such as an area law for the spatial Wilson loop and 
a mass gap \cite{GO,COOT,DJMN,HO,CORT,BISY,RTY,Min}. 
A very difficult unsolved problem in this approach is how to 
take the continuum limit, which corresponds to taking the curvature
of the background to infinity. 

A different proposal, due to Polyakov \cite{Poly1}, 
is to consider a string theory
with RR fields and a diagonal GSO projection that removes all fermions from
the spectrum while preserving modular invariance at one loop. Such theories
are referred to as type 0A or 0B, depending on the way one implements the 
projection \cite{DH,SW}. They have open string descendants that are
of interest for the D-brane construction \cite{BS,S1,S2,A,BG}. 
The bulk fields of these theories are the tachyon, the usual 
massless fields of the respective type II theories (with a doubling of the
RR part of the spectrum) and an infinite tower of bosonic massive modes.
If D-branes are included, the GSO projection in the open string sector
removes the world-volume tachyon and all the fermionic partners of the 
gauge bosons, leaving, in the critical case, 
only a bosonic Yang-Mills-Higgs theory obtained by 
dimensional reduction of pure Yang-Mills in $9+1$ dimensions to $p+1$
dimensions. For $p=3$ this theory is asymptotically free (recall that the 
additional presence of the fermions would make the theory exactly finite).

Klebanov and Tseytlin \cite{KT} have recently made the important 
observation that a large
background RR field provides a shift in the tachyon potential that effectively 
induces the tachyon to condense at a value of order one. 
The tachyon condensate has three effects: first it makes
the theory well defined in spite of the naive instability, second it induces 
an effective central charge proportional to $\langle T\rangle^2$ 
even in $d=10$, and 
third, it provides a mechanism for breaking conformal invariance in the 
world volume theory.

Assuming that the tachyon condenses, it pays off to consider the theory
off criticality, that is at arbitrary values of $d$, with an 
effective central charge given by
\beq
    \ce = 10-d + \frac{d-2}{16} \langle T\rangle^2. \label{ceff}
\eeq
This has the advantage of eliminating some or all
of the Higgs fields from the Yang-Mills theory under study, leaving in 
principle only the pure gauge theory if $d=p+2$. 
Also, if $d/2 - 2 < p \leq d - 2$ and $d\not = p+3$
there exist solutions to the Einstein equations having a constant dilaton, 
suggesting the existence of a conformal fixed point \cite{Poly1}.
Compactifying the conformal theory provides an alternative way to break
the conformal symmetry introducing a deformation parameter proportional
to the size of the compactified manifold.

In this paper we study the simplest example of such mechanism --- the
compactification on an Euclidean time circle of period inversely proportional 
to Hawking's temperature. In section two we write down the 
tree level gravity equations under the assumption that the tachyon 
condenses and that there is only one RR field present. 
We solve these equations and show that it is possible to have 
arbitrary temperature while keeping the dilaton constant.

In section three we compute the entropy for this solution and show that
it agrees with what is expected from a gas of weakly interacting YM particles.
We compute the entropy in two different ways, by evaluating the free energy 
(effective action)
and the horizon area, finding the same results. 
In the effective action approach
all the contribution comes from the RR fields.

In section four we
compute the spatial Wilson loop in the two limiting cases and show that  
it evolves from the behavior expected from a conformal field theory
($1/L$ potential) to that of a confining theory (area law). However,
compared with the standard construction, the bare string tension 
scales differently with the bare coupling\footnote{For a discussion of the 
Wilson loop in relation to zig-zag symmetry see \cite{AGO}.}. 
The fact that in this 
context one has one less free parameter than in the standard 
type II/M-theory construction presents a potentially serious problem in taking
the continuum limit. We also present the equation for the mass gap as a 
straightforward generalization of the previous analysis \cite{W2}. It is 
puzzling that the area law and mass gap persist for $p>4$. We comment on 
the possible physical meaning of such result.

\section{The gravity equations}

Assuming that the tachyon condenses, the one loop $\sigma$-model 
$\beta$-functions (tree level gravity equations) for the metric $g$, 
dilaton $\Phi$ and 
RR $p+1$ gauge potential $C$ can be derived from the action\footnote{
Throughout this paper we set $\alpha^\prime = 1$.}
\beq
    I=\int dx^d\sqrt{|g|}\left\{e^{-2\Phi}\left(\frac{\ce}{2}+R+
      4(\partial\Phi)^2\right) - \frac{1}{2(p+2)!} G^2\right\},
    \label{action} 
\eeq
where $\ce$ is given in (\ref{ceff}), $G=dC$ is the RR $p+2$ field strength
and $d$ includes the Liouville direction $r$. For us $r$ is just another
coordinate representing the radial direction.

The equations of motion are
\beqs
     R_{\mu\nu} &=& -2 D_\mu D_\nu \Phi + 
             e^{2\Phi} T_{\mu\nu}\label{einstein}\\
     \ce/2 + R  &=& 4 D_\mu \Phi D^\mu \Phi 
              - 4 D^\mu D_\mu \Phi\label{dilaton}\\
     D^\mu G_{\mu\indi} &=& 0, \label{gauge}
\eeqs
where 
\beq
    T_{\mu\nu} = \frac{1}{(p+1)!}\left(G_{\mu\indi}G_\nu^{\;\;\;\;\indi}
                 - \frac{1}{2(p+2)} g_{\mu\nu} G_{\indii}G^{\indii}\right)
     \label{stress}
\eeq
is the stress energy tensor of the RR field. Note that (\ref{stress}) is 
traceless only if $d=2p+4$.

We want to deform the solution found in \cite{Poly1} to allow for a non zero
temperature. We thus consider the following ansatz:
\beq
    g_{rr}=1;\;\;g_{00}=\gamma^2(r);\;\;g_{ij}=a^2(r)
    \delta_{ij};\;\;
    g_{ab}=b^2(r)\hat g_{ab},
\eeq
where $r$ is the Liouville direction, $0$ is the Euclidean time,
$i, j=1\cdots p$ the spatial coordinates and $a, b=1\cdots d-p-2$
the (possible) internal coordinates of a sphere. 
The Greek indices in (\ref{einstein},\ref{dilaton},\ref{gauge},\ref{stress}) 
are meant to run over the whole range $\mu = (r, 0, i, a)$
The metric $\hat g$ is normalized to have 
\beq
    \hat R_{abcd}=(\hat g_{ac} \hat g_{bd} - \hat g_{ad} \hat g_{bc}).
\eeq
With the further ansatz that 
\beq
    \Phi\equiv\Phi(r) \quad \hbox{and} \quad
     C_{0i_1\cdots i_p} = c(r)\epsilon_{i_1\cdots i_p},
\eeq
the equation of motion (\ref{gauge}) gives the standard solution
\beq
    c^\prime = N \frac{\gamma a^p}{b^{d-p-2}}
\eeq
that can be used to eliminate $c$ from the other 
equations. The remaining equations (\ref{einstein}, \ref{dilaton}) yield
\beqs
    \frac{\gamma^{''}}{\gamma}+p\frac{a^{''}}{a}+(d-p-2)\frac{b^{''}}{b}&=&
       2\Phi^{''} + \frac{N^2}{b^{2d-2p-4}}e^{2\Phi}\nonumber\\
    \frac{\gamma^{''}}{\gamma}+p\frac{\gamma^{'}}{\gamma}\frac{a^{'}}{a}
       +(d-p-2)\frac{\gamma^{'}}{\gamma}\frac{b^{'}}{b}&=& 
       2\frac{\gamma^{'}}{\gamma}\Phi^{'}+
       \frac{N^2}{b^{2d-2p-4}}e^{2\Phi}\nonumber\\
    \frac{a^{''}}{a}+\frac{\gamma^{'}}{\gamma}\frac{a^{'}}{a}+
       (p-1)\frac{a^{'2}}{a^2} +(d-p-2)\frac{a^{'}}{a}\frac{b^{'}}{b}&=&
       2\frac{a^{'}}{a}\Phi^{'}+\frac{N^2}{b^{2d-2p-4}}e^{2\Phi} 
       \label{ansa}\\
    \frac{b^{''}}{b}+\frac{\gamma^{'}}{\gamma}\frac{b^{'}}{b}+ 
       p\frac{a^{'}}{a}\frac{b^{'}}{b}+(d-p-3)\frac{b^{'2}}{b^2}-
       \frac{d-p-3}{b^2}&=&2\frac{b^{'}}{b}\Phi^{'}-
       \frac{N^2}{b^{2d-2p-4}}e^{2\Phi}\nonumber\\
     4\Phi^{'2}-2\Phi^{''}-2\left(\frac{\gamma^{'}}{\gamma}+ 
       p\frac{a^{'}}{a}+(d-p-2)\frac{b^{'}}{b}\right)\Phi^{'}&=&
       \frac{\ce}{2}-\frac{(2p+4-d)N^2}{b^{2d-2p-4}}e^{2\Phi}.\nonumber
\eeqs

The advantage of working in a non critical theory is that (\ref{ansa})
admit solutions with a constant dilaton even for $p\not = 3$. 
Such solutions are those of interest to us, so
let us specialize to this case by setting
$\lambda_{p+1}=N e^\Phi=$ const. 
($\lambda_{p+1}$ is the 't Hooft coupling in the appropriate units of 
$\alpha^\prime$).
We also search for solutions with constant $b$, so that the equations 
reduce to
\beqs
    \frac{\gamma^{''}}{\gamma}+p\frac{a^{''}}{a}&=&
       \frac{\lambda_{p+1}^2}{b^{2d-2p-4}}\nonumber\\
    \frac{\gamma^{''}}{\gamma}+p\frac{\gamma^{'}}{\gamma}\frac{a^{'}}{a}&=& 
       \frac{\lambda_{p+1}^2}{b^{2d-2p-4}}\nonumber\\
    \frac{a^{''}}{a}+\frac{\gamma^{'}}{\gamma}\frac{a^{'}}{a}+
       (p-1)\frac{a^{'2}}{a^2}&=&
       \frac{\lambda_{p+1}^2}{b^{2d-2p-4}} \label{consta} \\
       \frac{d-p-3}{b^2}&=&
       \frac{\lambda_{p+1}^2}{b^{2d-2p-4}}\nonumber\\
       \frac{\ce}{2}-\frac{(2p+4-d)\lambda_{p+1}^2}{b^{2d-2p-4}}&=&
       0.\nonumber
\eeqs

The equations don't allow for $d=p+3$, that is, for the 
compact dimension to be a circle.
When $d > p+3$, from the last two equations one can solve
for the 't Hooft coupling and for $b$, which are fixed to be (in units of 
$\alpha^\prime$)
\beqs
    \lambda_{p+1}^2&=&(d-p-3)\left(\frac{2(2p+4-d)(d-p-3)}{\ce}\right)^{d-p-3}
    \nonumber\\
    b^2&=&\frac{2(2p+4-d)(d-p-3)}{\ce}.
    \label{blambda}
\eeqs
The zero temperature solution is that of 
\cite{Poly1} and is found by setting
$a=\gamma=\exp(r/R)$, where $R$ is the radius of curvature. 
Eqs. (\ref{consta})
and (\ref{blambda}) then give
\beq
     R^2=\frac{2(p+1)(2p+4-d)}{\ce} \label{R}
\eeq
and the relation between the 't Hooft coupling and the radius 
of curvature reads 
\beq
     \lambda_{p+1}^2 \sim \left(R^2\right)^{d-p-3}.
     \label{scaling}
\eeq
Note that for $d=10$, $p=3$ we obtain Maldacena' s scaling. We
emphasize that it is only in the case of critical type II theory that 
one is truly
free to vary the parameters in (\ref{scaling}) although it is tempting to
hope that a better control of the tachyon field will allow to give a precise
physical meaning to the type 0 construction as well.

The case with no compact dimensions ($d=p+2$) should describe a theory without
any scalar. In this case the quantity $b$ drops out of the equations 
(\ref{ansa}), whose solution is then
\beq
    \lambda_{p+1}^2=\frac{p+1}{R^2}=\frac{\ce}{2(p+2)}.
\eeq  
Notice the peculiar behavior of $R$ that scales 
in a way inversely proportional
to the 't Hooft coupling, contrary to the standard situation. 
Again, this dependence should be interpreted with a grain of salt because
at this stage both values are fixed in terms of $\ce$ and, barring a novel
mechanism that allows to vary the effective central charge, we cannot take
the limit $\lambda_{p+1}\to 0$. 

The thermal deformation of this solution is more easily obtained by going
to the gauge
\beq
     ds^2 = \frac{\rho^2 f(\rho)}{R^2}dt^2 + \frac{\rho^2}{R^2}dx_i^2+
            \frac{R^2}{\rho^2 f(\rho)}d\rho^2 + b^2 d\Omega^2. \label{ga}
\eeq
Substituting into (\ref{consta}) yields
\beq
    (p+1) f(\rho) + \rho \frac{df}{d\rho} = p+1,
\eeq
whose solution is
\beq
     f(\rho) = 1 -  \frac{\rho_T^{p+1}}{\rho^{p+1}}.
\eeq
Notice that in the
extremal limit there exist solutions with $AdS_{p+2}\times S^{d-p-2}$ 
geometry for generic values of $p$ and $d$.
The Hawking temperature for this solution is easily computed to be
\beq
      T_H = \frac{p+1}{4\pi}\frac{\rho_T}{R^2}.
\eeq

\section{Thermodynamics of non-critical p-branes}

In this section we investigate the thermodynamic properties of the system
and find evidence in favor of the conjectured string theory/
gauge theory correspondence.
We compute the entropy \cite{HP,W2} of the non-critical, non-extremal
p-brane solution using two different methods and give
an interpretation of the results in terms of the light degrees 
of freedom living
on the brane, namely Yang-Mills theory \cite{GKP}. 
A similar analysis is performed in
\cite{BISY} for the type II D-branes.

For Yang-Mills theory in the weak coupling limit, one can neglect 
interactions 
between gluons and compute microscopically thermodynamic quantities using 
a free Bose gas approximation. In the case of $SU(N)$ 
gauge theory in $p+1$ dimensions
the energy and entropy per unit volume read
\beq
	\frac{E}{V} \sim N^2 T^{p+1}\qquad\qquad
	\frac{S}{V} \sim N^2 T^{p}
\eeq
which are, up to a numerical coefficient, dictated just by
dimensional arguments, $N^2$ being the number of degrees of freedom.

Let us follow \cite{W2} and \cite{HP} 
and identify the free energy $F$ of the black-brane as the 
(subtracted) Euclidean
action (c.f.r. (\ref{action})) times the Hawking temperature 
($\beta=\frac{1}{T_{H}}$),
\beq
	\beta F=I_E[g_{\mu\nu},\Phi,G;T_{H}]-
	I_E[g_{\mu\nu},\Phi,G;0]
\eeq
where as usual we subtract the zero temperature 
action to get a finite result.
Notice that by virtue of equation of motion 
(\ref{dilaton}) the Einstein and cosmological terms
drop out for constant dilaton and the action
gets contribution only from the RR field. In fact,
\beq
   I_E=\int dx^d \sqrt{|g|} \frac{1}{2(p+2)!} G^2=
   \frac{1}{2}\frac{N^2}{R^{2d-2p-4}} \int dx^d\sqrt{|g|}  
\eeq
After putting a cutoff in the ``radial'' integration, one has to evaluate two
invariant volumes, where in the black-brane configuration the integration is
to be performed in the physical region outside the horizon,
\beq
      V(\rho_{\infty})=\int^{\beta}_{0}dt \int^{\rho_{\infty}}_{\rho_{T}}d\rho
      \int d^px\frac{\rho^{p}}{R^{p}}\int R^{d-p-2}d\Omega_{d-p-2}
\eeq 
and
\beq
      V_{0}(\rho_{\infty})=\int^{\beta'}_{0}dt \int^{\rho_{\infty}}_{0}d\rho
      \int d^px\frac{\rho^{p}}{R^{p}}\int R^{d-p-2}d\Omega_{d-p-2}
\eeq 
and let $\rho_{\infty}\to\infty$ after subtracting them. Here the radius of
compactification $\beta'$ has to 
be matched with $\beta$ for the hyper-spheres in the two geometries to be
comparable, i.e.
\beq
     \beta'\frac{\rho_{\infty}}{R}=\beta\frac{\rho_{\infty}}{R}
     \sqrt{1-\frac{\rho_{T}^{p+1}}{\rho_{\infty}^{p+1}}}.
\eeq
The result is
\beq
 \beta F=\frac{N^2}{2R^{2d-2p-4}}\lim_{\rho_{\infty}\to\infty}(V-V_{0})\sim
  -R^{4+2p-d}\Omega_{d-p-2}N^2V_p\frac{1}{\beta^{p}}
\eeq
where $\Omega_{d-p-2}$ is the volume of the ($d-p-2$)-sphere and $V_p$ is the
total volume of the $p$-space.
Now recall from (\ref{R}) that $R\sim 1$ so that we finally get the
energy
 
\beq
E=\frac{\partial}{\partial \beta}(\beta F)\sim N^2V_pT_{H}^{p+1}
\eeq
and the entropy 
\beq
S=\beta (E-F)\sim N^2 V_p T_{H}^{p}. \label{entro}
\eeq

One can also compute the Bekenstein-Hawking entropy. In fact, going to the 
Einstein frame\footnote{This is the same as doing the calculation in the 
string frame and remembering that now the Newton constant depends on the 
dilaton and behaves like $\lambda_{p+1}^2/N^2$.}
\beq
ds^2_{E}=e^{-\frac{4}{d-2}\Phi}ds^2
\eeq
the area of the horizon is easily found to be (recall that $b\sim R$)
\beq
A\sim\Omega_{d-p-2}R^{4+2p-d} N^2V_{p}T_{H}^p,
\eeq
also in agreement with (\ref{entro}).
So we find that the Hawking
relation is reproduced and the entropy has the ideal gas scaling behavior.

This result is in agreement with the conclusions of \cite{KT2} for critical
black p-branes, where it is pointed out that constant dilaton is a sufficient
condition for such a scaling. Nevertheless, off criticality allows
for more general values of $p$.
  
Thus, we argue that there should be a correspondence between the gravity 
approximation of type 0 string theory and a non supersymmetric
Yang-Mills theory of $N^2$ degrees of freedom, 
finding support for Polyakov's conjecture.

\section{The spatial Wilson loop and the mass gap}

We are now in the position of computing the Wilson loop for the theory
described by the metric in section two using the techniques exposed in
\cite{SY,M2} and further developed in \cite{RTY,BISY}. 

Let us parameterize the world sheet
of the string as $x^1 = \sigma$, $x^2 = \tilde\sigma$, $\rho =\rho(\sigma)$
where $-L/2<\sigma<L/2$, $-\tilde L/2<\tilde\sigma<\tilde L/2$ and
$L << \tilde L$. The action to be minimized is
\beq
     S=\frac{\tilde L}{2\pi}\int_{-L/2}^{L/2} d\sigma \sqrt{
       \frac{\rho^{'2}}{1-\rho_T^{p+1}/\rho^{p+1}} + \frac{\rho^4}{R^4}}.
\eeq
The conserved quantity derived from this action is
\beq
    \frac{1}{\rho^4}\sqrt{\frac{\rho^{'2}}{1-\rho_T^{p+1}/\rho^{p+1}} + 
    \frac{\rho^4}{R^4}} = \frac{1}{\rho_0^2 R^2},
\eeq
where $\rho(0)= \rho_0$ and $\rho^{'}(0) = 0$ for symmetry reasons.
$\rho_0$ measures how close the world sheet approaches the horizon
at $\rho_T$ and the behavior of the Wilson loop is governed by the ratio
$\epsilon = \rho_T/\rho_0$. For $\epsilon\to 0$ we recover the conformal 
fixed point, whereas for $\epsilon\to 1$ we should approach the $p$ 
dimensional theory.

The minimum action is given by the integral (after subtracting the infinite 
energy of the string)
\beq
     S_{\mathrm min}=\frac{\tilde L \rho_0}{\pi}\left\{
     \epsilon -1 +\int_1^\infty dy\left[
     \frac{y^{(p+5)/2}}{\sqrt{(y^4-1)(y^{p+1}-\epsilon^{p+1})}} - 1
     \right] \right\}, \label{actionmin}
\eeq
where $\rho_0$ is expressed in terms of $R$, $L$ and $\rho_T$ by the
implicit function
\beq
     \frac{L}{2}=\frac{R^2}{\rho_0}\int_1^\infty dy\;
     \frac{y^{(p-3)/2}}{\sqrt{(y^4-1)(y^{p+1}-\epsilon^{p+1})}}. 
     \label{lenght}
\eeq
In the regime $\epsilon\to 0$ we obtain the results of \cite{SY,M2}:
$S_{\mathrm min}\sim R^2\times (\tilde L/L)$. Note that this is the 
same behavior as in \cite{SY,M2} only if expressed in terms of $R$. The 
relation between $R$ and the 't Hooft coupling 
being different (c.f.r. (\ref{scaling}))
if not in the critical dimension.

The interesting regime is when $\epsilon\to 1$. In this case, both integrals 
in (\ref{actionmin}), (\ref{lenght}) scale like $|\log(1-\epsilon)|$ and
we must eliminate the divergence by taking the ratio of the two quantities.
This leaves a dependence on $\rho_0$ but this is easily fixed by realizing
that as $\epsilon\to 1$, $\rho_0\to\rho_T$, yielding
\beq
    S_{\mathrm min}=\frac{\rho_T^2}{2\pi R^2} 
    \times \tilde L L. \label{arealaw}
\eeq
Eq. (\ref{arealaw}) represents the area law for the $p$ dimensional gauge 
theory, from which one can read off the bare string tension (always in units 
of $\alpha^\prime$)
\beq
    \tym=\frac{\rho_T^2}{R^2} \sim T_H^2. \label{weird}
\eeq
We immediately see a potentially serious problem with this construction. In
the most optimistic scenario, one would like to compute the renormalized 
string tension by taking the limit $T_H\to\infty$ while the $p$ dimensional
coupling $\hat\lambda_p$ goes to zero as $1/\log(T_H/\Lambda_{\mathrm QCD})$.
So far, this computation has been out of reach even for the standard 
construction. At least in that case, however, one has two truly independent
bare parameters to vary, namely $T_H$ and $\hat\lambda_p$. Here instead, 
the relation $\hat\lambda_p = \lambda_{p+1}T_H$ and the fact that 
$\lambda_{p+1}$ is fixed to be of order one by the equations of motion forces
$\hat\lambda_p \sim T_H$. We are thus led back to the issue raised in section
two on whether it is possible to relax eqs. (\ref{blambda}). This issue 
remains open.

Finally, we address the question of whether a mass gap will emerge in
the $p$ dimensional theory (at zero temperature), consistently with the
area law found above.

As explained in \cite{W1,W2} it will be sufficient to study the equation of
motion of a quantum field propagating in the background given by (\ref{ga})
and determine its spectrum in the $p$-dimensional sense. So, let us consider
the dilaton equation of motion (\ref{dilaton}). 
In spite of the presence of the cosmological constant, the 
constant dilaton background
renders the fluctuation field effectively 
massless (in $d$ dimensions), so that
we must still solve for
\beq
    \partial_\mu (\sqrt{|g|}g^{\mu\nu}\partial_\nu \delta\Phi)=0
    \label{eqdilaton}
\eeq
and we search for solution of the form $\delta\Phi=\chi(\rho)e^{ikx}$, 
$x\in {\mathrm\bf R}^p$.

After defining $y=\frac{\rho}{\rho_T}$ the equation of motion for $\chi$
following from (\ref{eqdilaton}) is
\beq
 \partial_y\left[(y^{p+2}-y)\partial_y\chi\right]+
 \rho_T^{-2}R^4M^2y^{p-2}\chi=0,
 \label{eqglueball}
\eeq
$M^2=-k^2$ being the mass squared of the glueball. $M\sim T_{H}$ as it should,
since the bare mass scales with the UV cutoff, the Hawking temperature in this
case.

Thus, it is straightforward to repeat the arguments of section
3.3 in \cite{W2} and conclude that the eigenvalue problem (\ref{eqglueball})
has normalizable solutions only for discrete and strictly positive values of
$M^2$. 
In fact (\ref{eqglueball}) actually reduces to the equation appearing 
in \cite{W2} for $p=3$, while for $p=4$ we also obtain a
mass gap for four dimensional gauge theory, in accordance with 
the area law. 

It is puzzling that we find a mass gap and an area law even for $p>4$.
This could be an artifact of the approximation but could also have a 
field theoretical explanation. Yang-Mills theory in more than four dimensions
is perturbatively non-renormalizable --- however, seen from the point of 
view of the $\epsilon$-expansion, the $4+\epsilon$ theory has a phase
transition at a finite value of the bare coupling constant. In the strong
coupling phase, the theory behaves, at low energies, in a way similar
to four dimensional Yang-Mills. It may happen that this is the phase 
relevant to the string theory description. 

\section{Acknowledgments}

We wish to thank R. Iengo, J. Kalkkinen, G. Mussardo and 
in particular J. Russo
for discussions. This work was supported in part by the European Union 
TMR program CT960012.


\begin{thebibliography}{99}

\bibitem{M1} J.~Maldacena, Adv. Theor. Math. Phys. {\bf 2} (1998) 231, 
{\tt hep-th/9711200}.
\bibitem{GKP1} S.S. Gubser, I.R. Klebanov, and A.M. Polyakov Phys. Lett.
{\bf B 428} (1998) 105, {\tt hep-th/9802109}.
\bibitem{W1} E.~Witten, Adv. Theor. Math. Phys. {\bf 2} (1998) 253, 
{\tt hep-th/9802150}.
\bibitem{W2} E.~Witten, Adv. Theor. Math. Phys. {\bf 2} (1998) 505, 
{\tt hep-th/9803131}.
\bibitem{GO} D.J. Gross and H. Ooguri,
Phys. Rev. {\bf D 58} (1998) 106002, \hfill\break {\tt hep-th/9805129}.
\bibitem{COOT} C. Csaki, H. Ooguri, Y. Oz and J. Terning,
{\tt hep-th/9806021}.
\bibitem{DJMN} R. De Mello Koch, A. Jevicki, M. Mihailescu and J. Nunes,
\hfill\break {\tt hep-th/9806125}.
\bibitem{HO} A. Hashimoto   and Y. Oz, {\tt hep-th/9809106}.
\bibitem{CORT} C. Csaki, Y. Oz, J. Russo  and J. Terning, {\tt hep-th/9810186}.
\bibitem{BISY} A. Brandhuber, N. Itzhaki, J. Sonnenschein, S. Yankielowicz 
J. High Energy Phys. {\bf 06} (1998) 001, {\tt hep-th/9803263}. 
\bibitem{RTY} S.J. Rey, S. Theisen and J. Yee, 
Nucl. Phys. {\bf B 527} (1998) 171, {\tt hep-th/9803135}. 
\bibitem{Min} J.A. Minahan, {\tt hep-th/9811156}. 
\bibitem{Poly1} A.M. Polyakov, {\tt hep-th/9809057}.
\bibitem{DH} L. Dixon and J. Harvey, Nucl. Phys. {\bf B 274} (1986) 93.
\bibitem{SW} N. Seiberg and E. Witten, Nucl. Phys. {\bf B 276} (1986) 272.
\bibitem{BS} M. Bianchi and  A. Sagnotti, Phys. Lett. {\bf B 247}  (1990) 517.
\bibitem{S1} A. Sagnotti, {\tt hep-th/9509080}.
\bibitem{S2} A. Sagnotti, Nucl. Phys. Proc. Suppl. {\bf B 56} (1997) 332, 
\hfill\break {\tt hep-th/9702093}. 
\bibitem{A} C. Angelantonj, {\tt  hep-th/9810214}.
\bibitem{BG} O. Bergman and M. Gaberdiel, Nucl. Phys. {\bf B 499} (1997) 183,
{\tt hep-th/9701137}.
\bibitem{KT} I.R. Klebanov and A.A. Tseytlin, {\tt hep-th/9811035}.
\bibitem{AGO} E. Alvarez, C. Gomez and T. Ortin, {\tt hep-th/9806075}. 
\bibitem{HP} S. Hawking and D. Page, Commun. Math. Phys. {\bf 87} (1983) 577.
\bibitem{GKP} S.S. Gubser, I.R. Klebanov and A.W. Peet, 
Phys. Rev. {\bf D 54} (1996) 3915, 
{\tt hep-th/9602135}.
\bibitem{KT2} I.R. Klebanov and A.A. Tseytlin,
Nucl. Phys. {\bf B 475} (1996) 164, {\tt hep-th/9604089}.
\bibitem{SY} S.J. Rey, J. Yee, {\tt hep-th/9803001}. 
\bibitem{M2} J. Maldacena, Phys. Rev. Lett. {\bf 80} (1998) 4859, 
{\tt hep-th/9803002}. 
\end{thebibliography}
\end{document}